\documentclass[%
 reprint,
superscriptaddress,
 amsmath,amssymb,
 aps,
]{revtex4-1}

\usepackage{graphicx}
\usepackage{dcolumn}
\usepackage{bm}
\usepackage[mathlines]{lineno}

\usepackage{braket}
\usepackage{amsmath}
\begin{document}

\preprint{APS/123-QED}

\title{Semiconductor quantum dots as an ideal source of polarization entangled photon pairs on-demand: a review}

\author{Daniel Huber}
\email{daniel.huber@jku.at}
\author{Marcus Reindl}
\author{Johannes Aberl}
\author{Armando Rastelli}
\email{armando.rastelli@jku.at}
\affiliation{Institute of Semiconductor and Solid State Physics, Johannes Kepler University, Linz, Altenbergerstr. 69, 4040, Austria}
\author{Rinaldo Trotta}
\email{rinaldo.trotta@uniroma1.it}
\affiliation{Institute of Semiconductor and Solid State Physics, Johannes Kepler University, Linz, Altenbergerstr. 69, 4040, Austria}
    \affiliation{Department of Physics, Sapienza University of Rome, Piazzale Aldo Moro 5, 00185 Rome, Italy}

\date{\today}

\begin{abstract}
More than 80 years passed since the first publication on entangled quantum states. In this period of time the concept of spookily interacting quantum states became an emerging field of science. After various experiments proving the existence of such non-classical states, visionary ideas were put forward to exploit entanglement in quantum information science and technology. These novel concepts have not yet come out of the experimental stage, mostly because of the lack of suitable, deterministic sources of entangled quantum states. Among many systems under investigation, semiconductor quantum dots are particularly appealing emitters of on-demand, single polarization-entangled photon-pairs. Although, it was originally believed that quantum dots must exhibit a limited degree of entanglement related to numerous decoherence effects present in the solid-state. Recent studies invalidated the premise of unavoidable entanglement degrading effects. We review the relevant experiments which have led to these important discoveries and discuss the remaining challenges for the anticipated quantum technologies. 

\end{abstract}

\pacs{Valid PACS appear here}
\maketitle

\section{Introduction}

\begin{figure}[htbp]
	\centering
		\includegraphics[width=90mm]{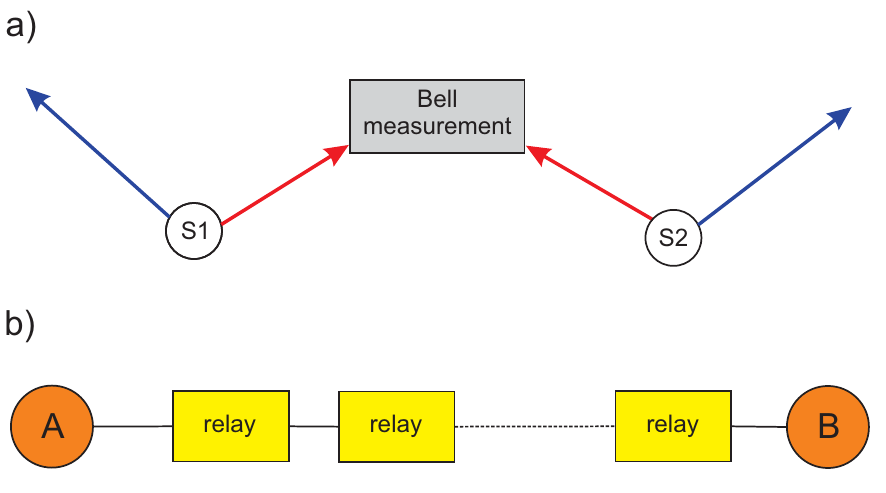}
	\caption{\textbf{Quantum repeater concept.} (a) In a quantum relay the photon source S1 and S2 emit pairs of entangled photons, where one photon of each pair is guided to the Bell state measurement apparatus enabling entanglement swapping. As a consequence, the outer photons remain in an entangled state. (b) In order to construct an entire quantum communication channel, several chained quantum relays have to be interfaced for a long-distance distribution of entangled states among individual parties A (Alice) and B (Bob).} 
	\label{fig:fig0}
\end{figure}

Entanglement is one of the most fascinating aspects of quantum mechanics. The idea, originally developed from a Gedankenexperiment (see Ref. \cite{PhysRev.47.777} and Ref. \cite{es1935}), took several decades to be experimentally realized, henceforward opened up a steadily growing field of research expanding to a manifold of sources generating "spookily" interacting quantum states. Further, innovative concepts of entanglement based computation and communication protocols were developed, leading to a dramatic increase of interest in the topic \cite{Zeilinger2017}. If we focus our discussion on the successful distribution of locally generated quantum states to any remote destination -  a fundamental requirement of quantum communication - photon states are the optimal solution due to their low environmental interaction. Nowadays, widely developed optical fiber networks could provide the communication basis of such photonic quantum states and finally pave the way towards envisioned concepts featuring secure communication via quantum cryptography \cite{RevModPhys.74.145} or even quantum computational networks \cite{Kimble:Nat2008}. However, this is a task far from being easy, as fiber channels are associated to decoherence and losses of transmitted states degrading the quantum information, thus crucially limits the communication distance \cite{RevModPhys.83.33,Pirandola2017}. A quantum state can not be amplified likewise to a classical communication channel and demands more advanced non-classical technology. Therefore, several concepts of so-called quantum repeaters have been developed among the past years \cite{ZollerNat2001,RevModPhys.83.33,PhysRevLett.98.190503,PhysRevLett.81.5932,PhysRevA.68.022301,Munro2012}. The very basic principle is shown in Fig. \ref{fig:fig0} (a), where two single photons emerging from two different entangled photon-pair sources are used to perform a Bell state measurement, eventually entangling the left-over photons. The procedure of entanglement swapping allows distant and independent photons, which have never interacted before, to be entangled. Ultimately, a chain-shaped arrangement of such quantum relays principally enables the distribution of entangled states over arbitrary long distances (see Fig. \ref{fig:fig0} (b)). We want to especially emphasize the substantial advantage to work with pairs of locally generated entangled photons-pairs in the subject of quantum repeater schemes, instead of the conditional and probabilistic NOON-state generation when relying on pure single photon sources \cite{Fattal2004}.  

The concept illustrated in Fig. 1 demands for sources of high quality entangled photons, fulfilling several requirements. In particular, each source has to emit single and entangled photons deterministically, with high purity, high efficiency, high indistinguishability and high degree of entanglement \cite{LU:2014NatPhoton}. In addition, it would be desirable to have sources to be integrable on chip-like infrastructures at competitive levels to state-of-the-art photonic technologies. In this regard, solid-state sources of entangled photons are of special interest \cite{Orieux2017}. In the past years the concept of parametric down conversion \cite{PhysRevLett.75.4337,Orieux2017} has been used for the majority of photon based entanglement experiments. However, besides the obvious advantages of the source, like its outstanding degree of entanglement and simplicity, one has to deal with severe drawbacks. First, a high degree of photon indistinguishability must be paid by a loss of source brightness. Second, the photon generation is a probabilistic process and follows Poissonian statistics, hence, there is a non-zero probability of generating multiple photon pairs during a single excitation cycle. Envisioning a quantum repeater scheme according to Ref. \cite{ZollerNat2001} it would be vulnerable to errors, thus lowering its overall efficiency \cite{Scarani2005}. It is rather clear that an alternative source is necessary. In 2000 O. Benson et al. presented the concept of a device using a single semiconductor quantum dot (QD) for the generation of polarization-entangled photon-pairs via the biexciton-exciton cascade \cite{PhysRevLett.84.2513}. QDs posses discrete quantum states similar to single atomic systems and therefore are also referred to as artificial atoms. In contrast to a single atom, QDs can be easily integrated within photonic chips and are compatible with actual semiconductor technologies. Recent research promoted QDs as emitters of single indistinguishable photons at high extraction efficiencies \cite{Senellart:NatPhoton2016,PhysRevLett.75.4337}. Nonetheless, it was commonly believed for a long time that QDs are not capable to reach a perfect degree of entanglement by cause of degradation effects related to the solid-state environment. In strong contrast, recent publications resolved the issue and revealed that QDs can be almost perfect emitters of polarization-entangled photon-pairs \cite{Fognini2017,Huber2018}.

Here, we aim to summarize the recent developments on QDs towards the perfect generation of entangled single photon states. In Sec. \ref{sec:entaglement} we introduce the entanglement generation process - the biexciton-exciton cascade - and discuss the main degrading effects altering the quantum superposition state. In Sec. \ref{sec:symQD} we provide information about the optimal growth properties of a QD, followed by promising device structures of energy tunable entangled photon sources in Sec. \ref{sec:energytuneable}. Further, we review in Sec. \ref{sec:excitation} the different excitation regimes focused on the optical population of a QD and its direct impact on the photon quality. In Sec. \ref{sec:perfSource} we present promising concepts of entanglement sources at high photon yield while simultaneously maintaining high indistinguishability, purity and ideal polarization entanglement.      

\section{Criteria for perfect entangled photons from a semiconductor quantum dot}\label{sec:entaglement}

\begin{figure}[htbp]
	\centering
		\includegraphics[width=90mm]{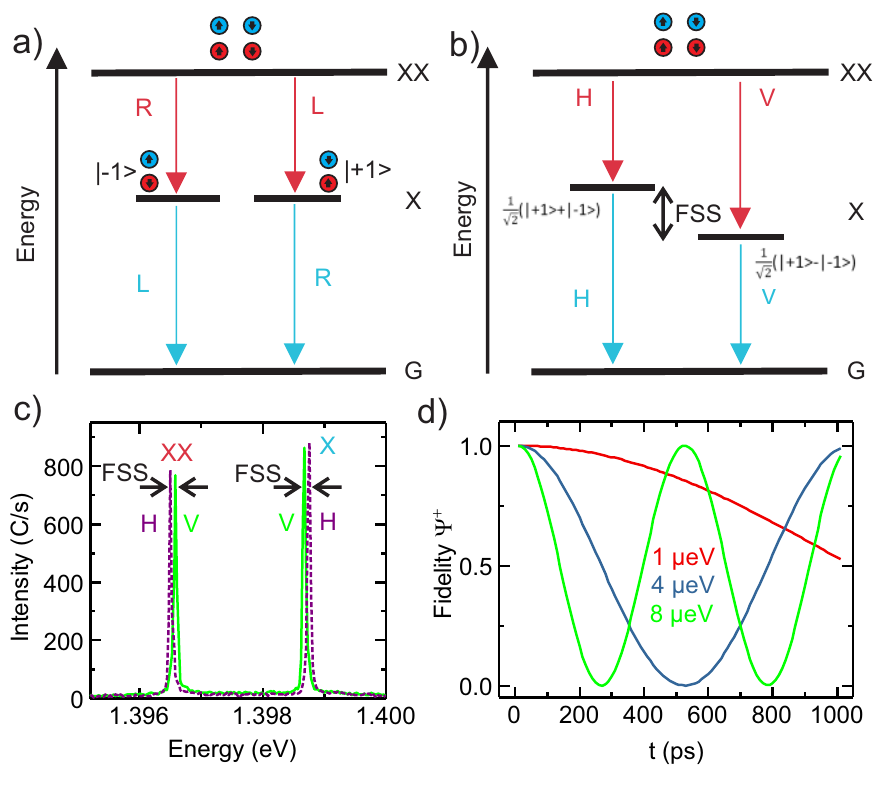}
	\caption{\textbf{Biexciton-exciton cascade and the effect of fine structure splitting.} (a) The biexciton state (XX) decays under the emission of a right circular (R) (left (L)) polarized photon to a single exciton state (X), which subsequently decays to the ground state (G) under the emission of an orthogonal-circular polarized photon. The polarization of the emitted photons is governed by the electron and hole spin configuration of the recombining electron-hole pair. (b) The anisotropic exchange interaction causes the intermediate exciton state to be energetically split by the so-called fine structure splitting (FSS). Consequently, a mixing of the exciton states appears, leading to the emission of two horizontally polarized photons (H) (vertically polarized photons (V)). (c) Polarization resolved $\mu$-photoluminescence measurement of a QD. The biexciton as well as the exciton are split due to the FSS. (d) Calculated fidelity to the entangled Bell state $\ket{\psi}$ as a function of the time delay between XX and X decay. In the calculation it is assumed that all other entanglement-reducing mechanisms but the FSS are absent. The principal effect of the FSS is a time-dependent oscillation of the fidelity as depicted for varying FSS of 1 $\mu eV$ (red), 4 $\mu eV$ (blue) and 8 $\mu eV$ (green).} 
	\label{fig:fig1}
\end{figure}

The fundamental process for the generation of polarization-entangled photon pairs in a QD is the so-called biexciton-exciton cascade as illustrated in Fig. \ref{fig:fig1} (a). Initially the QD is prepared in the biexciton state, which consists of two electron-hole pairs. This state decays via a recombination of one electron-hole pair under the emission of a single photon and leaves the quantum dot in a single exciton state, which subsequently decays via the emission of another single photon. The two electrons in the biexciton state have a spin quantum number of $m_{z}=+\frac{1}{2}$ and  $m_{z}=-\frac{1}{2}$, while the hole spins - in an ideal QD governed by heavy hole valance band states - have a spin of $m_{z}=+\frac{3}{2}$ and  $m_{z}=-\frac{3}{2}$. The polarization of the emitted photons is therefore determined by the total momentum of the recombining electron-hole pair, where only transitions with a change in the total momentum of $\pm 1$ are optically active. Hence, the cascade follows either the left or right decay channel - degenerated in energy - leading to a sequence of circular right (R) biexciton and circular left (L) exciton photons, or vice versa. The resulting two-photon state can be written as:

\begin{equation}
      \ket{\psi^{+}}=1/\sqrt{2}(\ket{L_{\text{XX}}}\ket{R_{\text{X}}}+\ket{R_{\text{XX}}}\ket{L_{\text{X}}}),
\label{eq:bellState}
\end{equation} 

where $L_{XX}$ ($R_{XX}$) and $L_{X}$ ($R_{X}$) are biexciton and exciton photons in the circular left (right) polarization base, respectively. This state corresponds to a maximally entangled Bell state. By using $\ket{H}=1/\sqrt{2}(\ket{L}+
\ket{R})$ and $\ket{V}= i/\sqrt{2}(\ket{R}-\ket{L})$ we can rewrite Eq. \ref{eq:bellState} in the linear horizontal (H) and vertical (V) base as:

\begin{equation}
      \ket{\psi}=1/\sqrt{2}(\ket{H_{\text{XX}}}\ket{H_{\text{X}}}+\ket{V_{\text{XX}}}\ket{V_{\text{X}}}).
			\label{eq:bellState2}
\end{equation} 

To take into account possible deviations of the real two-photon state from the ideal $\ket{\psi^{+}}$ state, a density operator formalism is commonly used. The density operator for a statistical ensemble of states is defined according to:

\begin{equation}
     \hat{\rho}=\sum_{i} p_{i}\ket{\phi_{i}}\bra{\phi_{i}},
			\label{eq:dmId}
\end{equation} 

where $p_{i}$ is the portion of the ensemble being in the state $\ket{\phi_{i}}$. In case of an ideal QD entangled photon emitter the density operator reduces to a pure state. Using the H and V states as basis, the matrix representation of the density operator is given by:

\begin{equation}
     \hat{\rho_{0}}=\ket{\psi}\bra{\psi}=\frac{1}{2}\begin{bmatrix}
1 & 0 & 0 & 1 \\
0 & 0 & 0 & 0 \\
0 & 0 & 0 & 0 \\
1 & 0 & 0 & 1
\end{bmatrix}.
			\label{eq:dm2}
\end{equation} 

The verification of photon entanglement emitted by a QD requires the reconstruction of its two-photon density matrix. This is done via polarization resolved cross-correlation measurements between biexciton and exciton photons emitted during the cascade decay. For a detailed description of the procedure and its evaluation we refer the interested reader to Ref. \cite{White:PRA2001}. The degree of entanglement can be quantified via different parameters namely concurrence, tangle, Peres criterion, negativity and fidelity (for a detailed summary see Ref. \cite{Orieux2017}). In works related to QDs the entanglement fidelity is a widely used figure of merit to compare different sources. It can be defined as:

\begin{equation}
     f^{+}=Tr[\rho_{mes}.\rho_{0}],
			\label{eq:fid}
\end{equation}

where $\rho_{mes}$ is the reconstructed density matrix obtained from the experiment and $\rho_{0}$ the density matrix of a pure Bell state according to Eq. \ref{eq:dm2}. If the measured density matrix is equal to the state in Eq.\ref{eq:bellState2} the resutling fidelity is one, while zero in the orthogonal case. Further, if the measured state is a mixed state, the fidelity is a measure of the probability to find an entangled photon pair in the state $\ket{\psi^{+}}$ in the ensemble. A classically correlated state yields a fidelity of $\frac{1}{2}$, while in the presence of entanglement $\frac{1}{2}<f^{+}\leq 1$ holds. We want to point out that the fidelity can also be measured with a reduced measurement set. A full reconstruction of the density matrix is not always required if the source is unpolarized and the measurement setup is not introducing any additional phase shifts. As a consequence the fidelity can be estimated as:

\begin{equation}
     f^{+}=\frac{1+C_{\text{linear}}+C_{\text{diagonal}}-C_{\text{circular}}}{4}.
			\label{eq:fidSimp}
\end{equation}
Here $C$ are the correlations visibilities according to 
\begin{equation}
      C_{\mu}=\frac{g^{2}_{XX,X}-g^{2}_{XX,\overline{X}}}{g^{2}_{XX,X}+g^{2}_{XX,\overline{X}}},
\end{equation}

where $g^{2}_{XX,X}$ and $g^{2}_{XX,\overline{X}}$ is the co- and the cross-polarized correlation measurement, respectively, in the polarization basis $\mu$. The measurement reduces to six cross-correlation measurements (see Ref. \cite{Huber2017,Keil2017}) instead of 16 for the full density matrix. 

In general a QD can underly various entanglement degrading effects, which drastically reduce the measured degree of entanglement and yield $f^{+}<1$. The dominant contributions are: (i) fine structure splitting (FSS), (ii) recapture processes, (iii) valence band mixing and (iv) exciton spin-flip processes. In the following, we would like to discuss these effects in more detail. 

(i) The FSS causes a mixing of the $\ket{\pm 1}$ spin states of the intermediate exciton states (compare \ref{fig:fig1} (a) and \ref{fig:fig1} (b)) due to a anisotropic electron-hole exchange interaction. In turn, the broken symmetry leads to a lifting of the degeneracy of the two excitonic states (see Fig. \ref{fig:fig1} (b)). As a result, the emission of a linear horizontal (H) (vertical (V)) polarized biexciton photon is followed by an equally polarized exciton photon during the cascade decay. The origin of the FSS can be illustrated by the exchange interaction Hamiltonian:

\begin{equation}
      H_{exch}=-\sum_{i=x,y,z}(a_{i}\hat{S}_{h,i}\cdot \hat{S}_{e,i}+b_{i} \hat{S}_{h,i}^3\cdot \hat{S}_{e,i}),
			\label{eq:hex}
\end{equation} 
    
where ${S}_{h,i}$ (${S}_{e,i}$) is the hole (electron) spin operator and $a_{i}$ and $b_{i}$ are the spin-spin coupling constants in the x,y,z spatial axis. The FSS arises for example, if the rotational in-plane symmetry of the wavefunction is reduced (for example due to a geometric symmetry lower than $D_{2D}$) leading to $b_{x}\neq b_{y}$ ($b_{x} = b_{y}$ for FSS=0). For a detailed description we refer the interested reader to Ref. \cite{Bayer:PRB2002} and Ref. \cite{Plumhof2012}. In contrast to the exciton, the biexciton does not show a FSS as the electrons and holes are in a singlet state. However, in polarization resolved $\mu$-photoluminescence measurements (see Fig. \ref{fig:fig1} (c)) one will observe the FSS in the exciton and biexciton spectral lines, simply due the fact that the biexciton decays into an exciton state.

Due to the FSS a time-dependent phase factor is introduced into the two-photon state. Therefore Eq. \ref{eq:bellState} transforms into:       

\begin{equation}
      \ket{\psi'}=1/\sqrt{2}(\ket{H_{\text{XX}}}\ket{H_{\text{X}}}+e^{\frac{i S t}{\hbar}}\ket{V_{\text{XX}}}\ket{V_{\text{X}}}),
			\label{eq:bellState2}
\end{equation} 

where $S$ is the FSS and $t$ the time span between the biexciton and exciton photon emission. The state is still entangled, but corresponds to $\ket{\psi^{+}}$ defined in Eq. \ref{eq:bellState2} only for decays with $t= \frac{2 N \pi \hbar}{S} = N T$ (with N an integer). In general $f^+$ will exhibit an oscillatory behavior as a function of decay time \cite{Stevenson:PRL2008}. It means that only a narrow subset of the decay events can be used as well-defined entanglement resource as shown in Fig. \ref{fig:fig1} (d) for different values of the FSS. If no event selection is applied, the time-averaged fidelity will be lower than unity. Assuming an exponential time distribution of the exciton photon emission, the resulting density matrix is given by:  

\begin{equation}
      \rho=\frac{1}{\tau_{1}}\int\limits_{0}^\tau e^{-t/\tau_{1}}\ket{\psi'}\bra{\psi'}dt,
      \label{eq:dmFSS}
\end{equation} 

where $\tau_{1}$ is the exciton lifetime and $\tau$ the integration time window. Hence, in case of a large FSS and/or long exciton lifetime one measures only classical correlations between the emitted photons for time windows $\tau \geq T$, where $T$ is the oscillation period of $f^{+}$. 

As a consequence, the first attempts towards the demonstration of entangled-photon generation with QDs did not show any evidence of non-classical behavior\cite{PhysRevB.66.045308}. In 2006 N. Akopian et al. confirmed for the first time polarization-entangled photons from a single QD \cite{PhysRevLett.96.130501}. The measurement relied on narrow spectral filtering of the biexciton and exciton photons to minimize the effect of the FSS. Similarly, one could also apply temporal filtering \cite{doi:10.1021/nl503581d,Stevenson:PRL2008,Ward2014,Salter2010,PhysRevApplied.8.024007}, such that photon pairs with high fidelity $f^{+}$ are selected. Implementing this scheme, Huwer et al. achieved in 2017 fidelities to the maximal entangled Bell state up to 0.92 \cite{PhysRevApplied.8.024007}. We want to emphasize that post-selection schemes suffer from severe photon-pair losses and inherently limits the possibility of on-demand generation. In particular temporal filtering introduces an unacceptable increase of complexity to any quantum optics system. A more favorable strategy is to directly eliminate the FSS itself. In Sec. \ref{sec:symQD} and Sec. \ref{sec:energytuneable} we present two successful strategies to eliminate the FSS without photon losses.    

(ii) In a recapture process (also referred to as re-excitation) the intermediate exciton level of the biexciton-exciton cascade is re-excited to the biexciton level before it decays to the ground state. The effect can be either optically driven (see Sec. \ref{sec:excitation}) or related to charged carriers trapped in the QD surrounding. An evidence of recapture is e.g. visible in the cross-correlation measurements between biexciton and exciton. It produces coincidence counts at negative time delays corresponding to exciton photons detected before the biexciton, a circumstance not expectable from the cascade`s temporal order \cite{10.1021/nl500968k}. The detected photons obviously feature no polarization correlation. The effect is well known (see Ref. \cite{Dousse:Nat2010,Kuroda:2013PRB,10.1021/nl500968k}) and particularly dominant under non-resonant excitation (see Sec. \ref{sec:excitation}). An additional scenario is that a cascade is followed by re-excitation within the same excitation cycle. If the XX of the first cascade and the X of the second cascade are detected while the other photons are lost to the detection inefficiency, the measured fidelity will be lowered. 

(iii) In an ideal system as described at the beginning of the chapter, the hole state of the exciton is purely heavy. The light hole band is considered to be energetically separated due to strong quantum confinement. In reality, the validity of this assumption is restricted to truly symmetric excitonic wave functions and an additional valance band mixing has to be considered in common QDs: The heavy-light hole mixing has two major effects: First, it contributes to the FSS and second, it leads to a depolarization of the exciton \cite{Lin2011,PhysRevB.87.075311}, with both effects degrading the measured degree of entanglement. Valence band mixing is not only related to the geometric in-plane symmetry of the QD, but also depends on strain anisotropies, interface intermixing effects and composition gradients within the QD. Even in absence of these perturbations significant valance band mixing can be introduced by a large vertical aspect-ratio of the wave function $\eta=l_{z}/l_{y}$ (where $l_z$ is the geometric parameter in growth direction and $l_{y}$ one in-plane component) as shown by Liao et al. in 2012 (see Ref. \cite{Liao2012}). Hence, a flat highly-symmetric strain-free QD shape is preferred as to avoid FSS and limit valance band mixing (see Sec. \ref{sec:symQD}).

(iv) Lastly, we want to discuss the effect of spin flips affecting the intermediated exciton decay. Such an event randomizes the polarization correlation between the biexciton and exciton photon. After a scattering event Eq. \ref{eq:bellState2} transforms to:

\begin{equation}
\ket{\psi_{s}}=1/2(\ket{H_{XX}H_{X}}+\ket{V_{XX}V_{X}}+\ket{H_{XX}V_{X}}+\ket{V_{XX}H_{X}}). 
			\label{eq:scattered_state}
\end{equation} 

If we first restrict ourselves to non-resonant excitation (see Sec. \ref{sec:excitation}) such a spin scattering can be explained by spin-flip processes due to the interaction with excess charges. The fluctuating electric field can induce an effective magnetic field altering the exciton spin via spin-orbit coupling \cite{MichlerQDs,Fognini2017}. 

Under (quasi) resonant excitation conditions, which means the absence of excess charges, spin scattering processes are still apparent. Another aspect of spin scattering is the interaction with the nuclear spins of the atoms forming the quantum dot and is completely decoupled from the excitation conditions \cite{MichlerQDs,Chekhovich2011,PhysRevLett.99.266802}. Thereby, it is assumed that the confined electron of the exciton is interacting via Fermi-contact interaction with the nuclear spins, while the heavy-hole dephasing is related to a dipole-dipole interaction \cite{RevModPhys.85.79}. In 2017 Fognini et al. performed high resolution (30 ps) time-resolved correlation measurements on wurtzite InAsP QDs in a tapered InP nanowire under quasi-resonant excitation (see Sec. \ref{sec:excitation}). Despite the presence of high nuclear spin In atoms in the QD and a long exciton life time (1\,ns) - implying a large impact of the nulcei induced spin dephasing - the authors show that these QDs are dephasing free and the model of spin scattering \cite{PhysRevLett.99.266802} is experimentally not supported. Further, exciton spin flips have not been observed in InAs/GaAs QDs on relevant time scales ($<30$ ns) at cryogenic temperatures ($<20$ K) \cite{PhysRevB.70.033301}.

In the work of Huber et. al. (see Ref. \cite{Huber2018}) the authors performed a measurement of the dependence of the entanglement fidelity versus the FSS under resonant excitation on a single GaAs QD. The results indicate the deviation from perfect entanglement compatible with residual spin scattering process with an unknown origin.

\subsection{Growth of symmetric quantum dots}\label{sec:symQD}

In the previous part we discussed that an anisotropy of the confinement potential leads to a finite FSS and valance band mixing. Over the last decade the majority of entanglement experiments have been performed on Stranski-Krastanow InGaAs QDs. Specifically, this material system suffers from anisotropic strain fields related to the growth mode, interface diffusion processes \cite{PhysRevB.64.165306} and large values of valance band mixing ($0.2<|\beta|<0.7$, where $\beta=0$ corresponds to a pure heavy hole and absence of valance band mixing \cite{Belhadj2010}). As a consequence, the probability of finding InGaAs QDs with FSS smaller than 1 $\mu$eV is of the order of a few per cent \cite{PhysRevB.89.205312}. In 2013 Juska et al. presented a work\cite{Juska2013} on  pyramidal site-controlled InGaAs$_{1-\delta}$N$_{\delta}$ QDs, which are grown on a patterned (111)B-oriented GaAs substrate via metal–organic vapour phase epitaxy. The growth technique leads to ensemble of QDs with high structural symmetry (see also Ref. \cite{Treu2012,Juska2015,Chung2016}) and up to 15$\%$ of the QDs can generate polarization-entangled photons with a  fidelity up to 0.72(4) under non-resonant excitation. A different attempt of symmetrically engineered excitonic wavefunctions based on the InAs material system is relying on the growth of large low-strain In$_{0.3}$Ga$_{0.7}$As QDs \cite{PhysRevB.90.045430} exhibiting fine structure splittings well below 10 $\mu$eV on the average, despite structural anisotropies. Here, the shallow confinement potential lowers the sensitivity of the wavefunction with respect to structural asymmetries of the QD. Details of the QD shape practically become negligible compared to the large wavefunction extension. Other material systems, except GaAs QDs which are introduced in more detail, have not yet met the demanding requirements of a high, and statistically frequent, excitonic confinement symmetry \cite{omfss1,omfss2,omfss3}. The results clearly highlighted that the shape of the QDs and the in-plane anisotropy of the confining potential have a dominant role in the underlying physics of the FSS. However, even in the case of high wavefunction symmetry, the InGaAs material system still suffers from the In-related large Overhauser field \cite{RevModPhys.85.79}, which leads to a fluctuation of the FSS over time \cite{Burk:arxiv2015}. From this point of view, a promising material system, namely GaAs/AlGaAs QDs, is of increasing interest. The material combination is characterized by a negligible lattice mismatch, which facilitates the realization of QDs with high in-plane symmetry combined with the small nuclear spin of Ga (3/2) compared to In (9/2), thus minimizing the effective Overhauser field perturbing the FSS \cite{Stevenson2013}.

We want to briefly discuss three approaches, which lead to as-grown GaAs QDs emitting highly entangled photons. The first method is a droplet epitaxy method \cite{Mano2000,Sanguinetti2003,Mano2006}, where pure Ga is evaporated on an AlGaAs surface, via molecular beam epitaxy. The Ga is forming droplets on the surface, which are subsequently crystallized by an additional As flux. After annealing, the QDs are capped with an AlGaAs layer forming the top barrier of the QD. In 2013 Kuroda et al. used such highly symmetric QDs to measure a fidelity of 0.86 (the exact FSS is not given) without the aid of photon post-selection. A drawback of the growth method is an unfavorable broad size distribution due to the low crystallization temperature used \cite{Heyn2009}. In 2017 Basso Basset et al. offered an alternative solution by presenting a modified approach on GaAs (111) substrates, where the Ga droplets are crystallized and the subsequent barrier layer is deposited at a high substrate
temperature of around 520$^\circ$C \cite{Basset2017}. The authors claim an improved crystallinity of the QDs by reducing the defect concentration typical for low-temperature growth eventually leading to a higher optical quality of the QDs. 

The second method is the fabrication of QDs by etching of nanoholes into a AlGaAs substrate via local droplet-etching \cite{Wang2007}. The nanoholes are subsequently filled with GaAs and capped, after annealing, with AlGaAs \cite{Heyn2009}. In 2013 Huo et al. demonstrated that, under optimized growth conditions, droplet-etched QDs yield an ultra small FSS with an average value of 4(2) $\mu$eV \cite{Huo:APL2013}. Further, a small valance band mixing ($<5\%$) is reported \cite{Huo:NatPhys,Huo2017}. In addition to the in-plane symmetry, a tall morphology is useful to reduce the FSS, as the exchange interaction depends on the strength of the electron-hole wave functions overlap \cite{PhysRevB.90.041304}. 

In 2017 Huber et al. measured the degree of entanglement and achieved a fidelity of 0.94(1), already at a finite FSS of 1.2(5) $\mu eV$, without any post-selection and is up to date the highest value of fidelity measured on a QD without any post-growth techniques. Similar values of entanglement were reported by Keil et al. - where the authors claim a fidelity of 0.91 at a FSS of 2.3 $\mu eV$ - using the same type of QDs \cite{Keil2017}. Both works rely on a resonant two-photon excitation scheme to directly populate the biexciton state (see Sec. \ref{sec:excitation}). 

In spite of the impressive progress towards the fabrication of entanglement-ready QDs \cite{PhysRevApplied.8.024007,Huber2017,Keil2017,Kuroda:2013PRB}, finding QDs with sub-$\mu$eV-FSS in an ensemble is still very challenging due to unavoidable fluctuations occurring during the high-temperature growth of such nanostructures. The mentioned experiments yielding record fidelities are always based on the search of "hero dots" with low FSS. Nonetheless, fidelities on the order of 0.94 are definitely an attractive value for ready-to-use quantum relays including error correction protocols \cite{PhysRevA.66.060302}.

\subsection{Energy tunable source of polarization entangled photons}\label{sec:energytuneable}

\begin{figure*}[htbp]
	\centering
		\includegraphics[width=180mm]{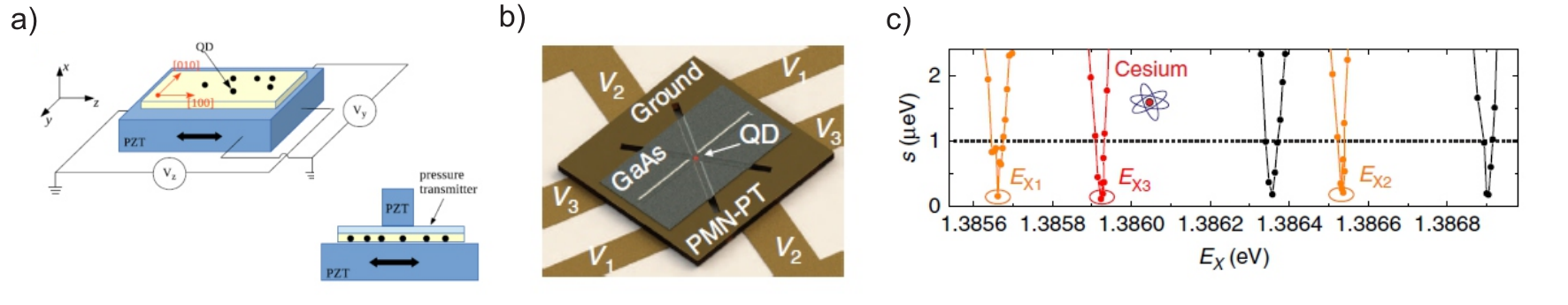}
	\caption{\textbf{Energy-tuneable entangled photon sources based on QDs.} (a) Device structure according to Wang et al. adapted from Ref. \cite{PhysRevLett.115.067401}. The piezoelectric actuator (PZT) can be used to apply two independent in-plane stresses via the voltages $V_{z}$ and $V_{y}$ to cancel the fine structure splitting. A second stressor on top of a transparent stress transmitter would allow for energy shifting. (b) Device structure according to Trotta et al. adapted from Ref. \cite{Trotta:2016NatCom}. Six pairwise connected piezoelectric legs on a micromachined PMN-PT substrate are used to apply any arbitrary in-plane stress configuration on the membrane via the voltages $V_{1},V_{2},V_{3}$ to erase the fine structure splitting and simultaneously tune the exciton (biexciton) emission energy. (c) Excitonic degeneracy tuning of the same, single QD at arbitrarily chosen emission energies. The figure is adapted from Ref. \cite{Trotta:2016NatCom}. The piezoelectric actuator from (b) allows to tune the fine structure to zero independently from the targeted emission energy. The minimum of the red curve coincides with the D2 line of a cloud of Cs atoms.} 
	\label{fig:fig2}
\end{figure*}

The compensation of residual asymmetries leading to a finite FSS can possibly resorted to external perturbations, such as the optical Stark effect, electric-, magnetic fields as well as anisotropic strain fields or a combination of them (for details see Ref. \cite{Plumhof2012,martin2017strain,PhysRevLett.109.147401,10.1021/nl500968k}).

Envisioning future application of QDs in the field of quantum communication one has to deal with an additional problem. A quantum relay as shown in Fig. \ref{fig:fig0} (a) relies on perfect indistinguishablility of the emitted photon pairs. It inherently demands perfect energy matching of photons emitted by different sources, basically requiring all QDs emitting fully identical entangled photons. As the wavelengths of QDs statistically differ from QD to QD the probability to find at least two QDs under this conditions - at zero FSS and equal energy - is $\approx 10^{-9}$ \cite{PhysRevLett.114.150502} in standard Stranski–Krastanow QDs. Furthermore, the entangled photons are ideally interfaced with other quantum systems like atomic clouds or N-V-centers (acting as quantum memories) adding another complex parameter to be controlled \cite{northup2014quantum}. Arguably, the desired condition is not achievable with realistic growth processes. It was shown that the emission energy of a QD can be precisely engineered using the mentioned external perturbations  \cite{Plumhof2012,martin2017strain}, but sophisticated applications demand for a device structure enabling independent control of emission energy and FSS. Only a device structure featuring simultaneous control of emission energy and FSS on any emitter of investigation, a truly scalable QD-based quantum communication network is feasible.

Since the commonly used perturbation fields are of vectorial (e.g. electric field) or tensorial (strain field) nature and each of them produces a distinct effect on the electronic structure of QDs, these perturbations provide in principle a large enough set of independent "tuning knobs" to control emission energy of X and XX photons as well as FSS. Two individual concepts were presented in 2015 by Wang et al. (see Ref. \cite{PhysRevLett.115.067401}) and by Trotta et al. (see Ref. \cite{PhysRevLett.114.150502}), both relying on properly engineered strain fields. The first concept (see Fig. \ref{fig:fig2} (a) ) uses a piezoelectric lead zirconic titanate ceramic stack with QDs glued on top. By applying two independent voltages the QDs are exposed to two independent in-plane strain components, which can be used to erase the FSS of an arbitrary QD \cite{doi:10.1063/1.4745188}. For energy tuning, a second independent stressor is placed on top of the sample. The extraction of the photons, however, require the technologically challenging use of a transparent strain transmitters between the sample and the top stressor.

The second concept, as shown in Fig.\ref{fig:fig2} (b) uses a micromachined piezoelectric structures ([Pb(Mg$_{1/3}$Nb$_{2/3}$)O$_{3}$]$_{0.72}$-[PbTiO$_{3}$]$_{0.28}$), with three stress fields acting on the QDs by three pairs of "legs", as shown in Fig. \ref{fig:fig2} (b). Three independent voltages ($V_{1}$,$V_{2}$,$V_{3}$) applied across the legs and the top (grounded) contact allow for full control over the in-plane stress state in a nanomembrane - hosting the QDs - bonded on top of the piezoelectric actuator \cite{sanchez:AM2016}. As a consequence, one can use two legs of the device to fully erase the FSS, while the third leg enables energy tuning at fixed FSS. As the membrane is open to the top, the emitted photons can be easily collected by microscope objectives. Furthermore, the structure enables the possibility to use QDs embedded in a diode structure as to electrically pump the entangled photon source (see Ref. \cite{Salter2010}). 

In 2016 Trotta et al. presented an experimental proof of their concept \cite{Trotta:2016NatCom}. The authors erased the FSS of an InAs QD and verified photon entanglement (with resulting fidelity of 0.80(5)) at different energies of the X transition. Additionally, they interfaced the exciton photons with the D1 line of cesium vapor and at the same time keep the FSS constantly below the spectral resolution of the setup ($<0.2$ $\mu$eV) (see Fig. \ref{fig:fig2}) (c). The QD-Cs interface allowed the authors to observe slow light\cite{10.1038/nphoton.2011.16}, and in particular slow entangled photons.

Other approaches enabling wavelength tunable sources of entangled photons with InAs QDs were presented in 2016 by Chen et al. using patterned thin film piezoelectric structures fabricated throughout focused ion beam and wet chemical etching\cite{Chen2016}. In the same year, Zhang et al. presented an approach using a monolithic piezoelectric actuator in combination with a vertical electric field\cite{10.1021/acs.nanolett.6b04539}. We want to emphasize that such "two-knob" devices act as wavelength tunable sources of entangled photons only if the structural anisotropy of the QD is oriented along the main stress direction of the piezoelectric actuator, which is given for around $30\%$ of the QDs according to Ref. \cite{Zhang2015}. The same consideration applies to approaches combining electric and magnetic fields with fixed orientation \cite{PhysRevApplied.1.024002}.  

In 2018 Huber et al. combined the patterned piezo structure with $\mu$-membranes hosting droplet-etched GaAs/AlGaAs QDs in a planar distributed Bragg reflector cavity \cite{Huber2018}. The authors achieved an entanglement fidelity up to 0.978(5) after taking into account setup-related entanglement deteriorating effects (phase shifts induced by the optical elements and laser background). Accordingly, the only known limit is a spin flip process with a characteristic scattering time of 14(10) ns. The origin of the spin scattering is not clear yet. A forced reduction of the excitonic emission lifetime from around 290 ps to 100 ps via Purcell enhancement \cite{Dousse:Nat2010} would allow GaAs QDs to reach levels of near-unity entanglement compatible to parametric down conversion sources with the mentioned benefit of deterministic operation.  
     
\subsection{The role of excitation}\label{sec:excitation}

\begin{figure*}[htbp]
	\centering
		\includegraphics[width=180mm]{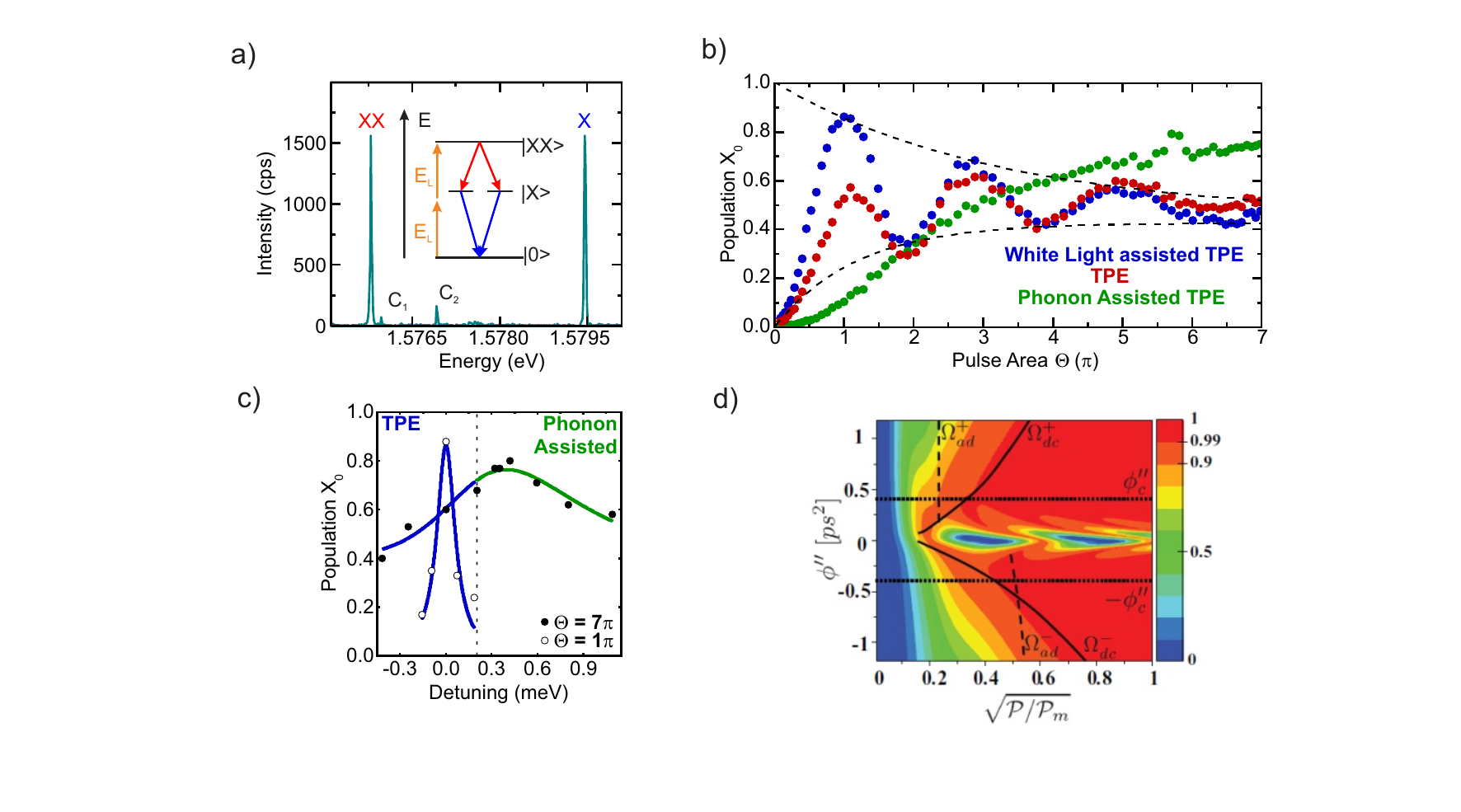}
	\caption{\textbf{Different coherent excitation schemes.} (a) Resonantly excited GaAs quantum dot adapted from Ref. \cite{Huber2017}. The exciton (X) and biexciton (XX) of similar intensity are visible. Two weaker, charged states appear (C1 and C2). Inset: For resonant two-photon excitation, the laser energy $E_{L}$ is tuned to half the energy of the biexciton state. Two photons are absorbed to directly populate the biexciton state. (b) Exciton ($X_{0}$) population under resonant two-photon excitation (TPE) (red), white light assisted TPE (blue) and phonon-assisted excitation (green) versus pulse area on the same QD. The envelope of the Rabi oscillation is modeled with a single exponential damping. As a consequence of the cascade process, the exciton exhibits the same population evolution as the resonantly driven biexciton (see Ref. \cite{Reindl:nano2017}). The figure is adapted from Ref. \cite{Reindl:nano2017}. (c)  Population of the X state as a function of the laser detuning for varying excitation power. While the traditional TPE (blue)
suffers from a steep drop in inversion efficiency, a stable plateau can be observed exploiting the QD phonon sideband (green). The figure is adapted from Ref. \cite{Reindl:nano2017}. (d) Biexciton population versus power and chirp parameters as to identify the adiabatic rapid passage excitation condition. The figure is adapted from Ref. \cite{PhysRevB.88.201305}.} 
	\label{fig:fig3}
\end{figure*}

The population of an excited state in a QD can be achieved in various ways. If we restrict our discussion to pulsed optical pumping one has three major options: (i) non-resonant excitation, (ii) quasi-resonant excitation and (iii) resonant excitation, which we will discuss in more detail:

(i) In non-resonant excitation the laser is mainly exciting carriers in the barrier material or in the wetting layer of the QD (if available) to photo-generate electron-hole pairs. Subsequently, these electrons and holes are captured by the QDs and relax to the lowest energy levels, the so-called s-shell. Arguably, the excitation regime is not favorable for an entangled photon generation supporting the appearance of pronounced recapture processes \cite{Dousse:Nat2010,Kuroda:2013PRB,10.1021/nl500968k}. Additionally, the fidelity is lowered by spin dephasing processes arising from exciton and excess charge interactions \cite{Fognini2017}. Besides the negative effects on the entanglement fidelity additional drawback are, first, the possibility to achieve deterministic population of the biexciton state is inhibited \cite{Jayakumar2013,Michler:NaturePhoton2014}, and second, the indistinguishability of the emitted photons suffers from the carrier relaxation induced time jitter and fluctuating electric fields \cite{Huber2015,Senellart:NatPhoton2016} (see Sec. \ref{sec:purity}).

(ii) In  contrast to non-resonant excitation, quasi-resonant excitation is a more favorable approach in generating entangled photon pairs. In this regime the electron-hole pairs are generated via excitation in a higher QD shell, e.g. p-shell, and the carriers normally undergo a fast relaxation process to the s-shell. The excitation allows for a "near" coherent population once resonantly driving the p-shell \cite{MichlerQDs}. In 2017 Fognini et al. studied the entanglement properties in a InAsP QD under quasi-resonant excitation, where no adverse effects related to the excitation could be identified \cite{Fognini2017}. However, in a recent study by Kir\v{s}ansk\v{e} et al. the authors showed that unde quasi-resonant excitation conditions the indistinguishability is essentially limited by the time jitter introduced via p-shell to s-shell relaxation processes \cite{Kirsansk2017}. 

(iii) In resonant excitation the electron-hole pairs are directly created in the s-shell of the QD. The resonant population of the biexciton state requires a resonant two-photon absorption process due to dipole selection rules \cite{PhysRevB.73.125304,PhysRevLett.73.1138}. The concept is shown in the inset of Fig. \ref{fig:fig3} (a), where the laser is tuned between the biexciton and exciton spectral line. The energy spacing - governed by the biexciton binding energy - is small (typically around 1-4 meV for InAs and GaAs QDs, respectively \cite{Michler:NaturePhoton2014,Huber2017,Keil2017}), such that the laser is energetically close to, albeit not overlapped, with the QD photons. A sophisticated rejection of the scattered laser light is unavoidable, as already minor contributions of the laser can significantly lower the single photon purity, indistinguishability and the degree of entanglement \cite{Huber2017,Huber2018}. It is particularly relevant once measuring the degree of entanglement of the source, as common techniques relying on cross-polarization between excitation and detection cannot be adapted \cite{Jayakumar2013}. According to current research, however, the resonant and quasi-resonant excitation do not show appreciable difference on the degree of entanglement, but only on the indistinguishability of the emitted photons. In fact, studies done by Ding et al. and Somaschi et al. proved that QDs are only exhibiting near-perfect degree of photon indistinguishability under resonant conditions (indistinguishabilities of 0.996(5) \cite{Senellart:NatPhoton2016} and 0.985(4) \cite{PhysRevLett.116.020401} are reported). Further, the resonant excitation allows for almost near unity population probability (Fig. \ref{fig:fig3} (b)) of the biexciton state after application of a single $\pi$-pulse \cite{PhysRevB.91.161302,Michler:NaturePhoton2014,Reindl:nano2017} - a condition of high importance in terms of quantum efficiency of the entangled photon source. Hence, to simultaneously optimize the degree of entanglement, photon indistinguishability, single photon purity and on-demand generation, a resonant excitation regime is obligatory \cite{Huber2017}. The only disadvantage is the severe QD-dependent sensitivity of the resonant condition. Small fluctuations in the laser pulse area, energy and QD environment result in a strong variation of the excited state population probability (see Fig.\ref{fig:fig3} (c)). Therefore, two alternative excitation schemes based on resonant excitation are introduced, as discussed below:

First, we would like to introduce the adiabatic rapid passage (ARP), an excitation technique well known for single atomic systems. Here, the resonant condition of the QD two-level system is established utilizing a chirped laser pulse, in contrast to the commonly used transform-limited $\pi$-pulse. The duration of the frequency-swept pulse should be significantly shorter than the spontaneous emission time ("rapid") but still slow enough in comparison to the Rabi frequency such that the excited state can be tracked adiabatically. The result of the light-matter interaction Hamiltonian is a traversed anti-crossing of the two possible eigenstates \cite{Hopkinson2011} causing the system to switch into the excited state at sufficient excitation power. The robustness of the scheme is theoretically demonstrated in Fig. \ref{fig:fig3} (d) and unambiguously shows the pulse area independent maximum state inversion as a function of the pulse chirp, strikingly different from the very sensitive Rabi oscillations of transform-limited excitation conditions. Nonetheless, the quantum dynamics of the ARP strongly relies on a complex control of the excitation conditions as well as minimized dephasing mechanisms of the coherent superposition of the two-level system such as carrier-phonon interactions \cite{Glassl2013b}.

An alternative excitation scheme which combines the simplicity of strictly resonant excitation with the robustness of the ARP is the so called phonon-assisted two-photon excitation. It was proposed \cite{Glassl2013} to tune a transform-limited, two-photon excitation laser slightly above the biexciton energy as to dress a vibrational quasicontinuum and, in the following, on-demand prepare the biexciton state despite the presence of strong carrier-phonon interactions. These interaction terms are the main obstacle to achieve high preparation fidelity, particularly pronounced in semiconductor QDs \cite{Forstner2003,Ramsay2010}. The proposal was verified in several works \cite{Ardelt.Hanschke.ea:2014,Quilter.Brash.ea:2015,Bounouar.Muller.ea:2015} and its comparison to standard resonant schemes is depicted in Fig. \ref{fig:fig3} (b). While traditional Rabi oscillations suffer from strong phonon-induced damping for increasing driving strength, the two-photon phonon-assisted excitation unfolds its real potential. At high excitation powers it is not only possible to prepare the QD with comparable levels of state inversion, but additionally, the excited state population is independent of laser intensity fluctuations. Most importantly, the phonon sideband dressing of a QD biexciton state is allowed for a certain range of laser energies (Fig. \ref{fig:fig3} (c)) and presents a feasible way to address complex systems of dissimilar QDs at the same time by stable, frequency-locked laser systems, a condition of high practical advantage impossible to realize with strict resonant excitation. Careful studies revealed that all advantages of $\pi$-pulse driven systems regarding their emission quality, in particular polarization entanglement, hold up for the two-photon phonon-assisted condition due to the negligible time-jitter introduced by the ultra-fast phonon relaxation times \cite{Reindl:nano2017}.

Besides of optical excited QDs, electrically driven sources of entangled photons are of special interest as the absence of an excitation laser would drastically reduce the complexity of the system \cite{Salter2010,PhysRevLett.108.040503}. QDs can be embedded into light-emitting diode structures \cite{Yuan102} and several approaches concerning pulsed entangled-light-emitting-diodes have been presented \cite{Zhang2015,Chung2016}. The highest fidelity under pulsed electrical pumping was achieved by Chung et al. in 2016 featuring $f^{+}=0.68(2)$ at a FSS of 0.2(2) $\mu$eV (see Ref. \cite{Chung2016}) without any post-selection. The results are significantly lower with respect to optical excitation \cite{Huber2017,Keil2017} and suggests entanglement degrading effects, apart from the FSS, are strikingly dominant (especially re-population \cite{Chung2016}). The full potential of these devices might only be accessible via an electrically driven resonant population process.

\section{Towards a perfect entangled photon source}\label{sec:perfSource}

QD-based sources already approach the state-of-the-art set by parametric down-converters \cite{shalm2015strong} in terms of near-unity entangled-state fidelities. However, besides a high degree of entanglement and the capability of generating photons at predefined energies, additional criteria must be met by the "perfect" source. For truly on-demand operation, one and only one photon pair should be emitted after each excitation cycle and, in addition, the photons emitted in subsequent cascades should be indistinguishable in all degrees of freedom apart from their polarization. Among these criteria, scalable applications are only feasible if near-unity fractions of the generated photons are extracted from the semiconductor matrix. 

\subsection{Single photon purity and photon indistinguishability}
\label{sec:purity}

The single photon purity of a photon source is characterized via a Hanbury Brown and Twiss (HBT) experiment. It allows to extract the second-order autocorrelation function $g^{(2)}(\tau)$ versus the time delay $\tau$. A perfect single photon source gives a $g^{(2)}(0)=0$ and a two-photon source gives $g^{(2)}(0)=0.5$ \cite{MichlerQDs}. In view of sources of entangled photon pairs, unprecedented values of purity alongside high entanglement fidelity have been achieved, mainly relying on the resonant two-photon excitation of droplet-etched GaAs QDs \cite{Huber2017,Keil2017}.

\begin{figure}[htbp]
	\centering
		\includegraphics[width=90mm]{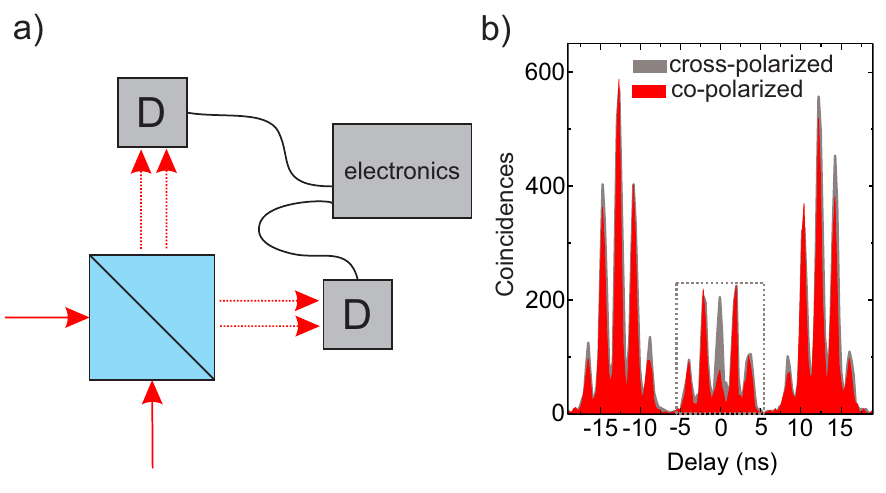}
	\caption{\textbf{Two-photon interference.} (a) Two photons impinge on a beam splitter and experience the Hong-Ou-Mandel effect, where indistinguishable photons always leave through the same output port. The outputs are equipped with detectors (D) to measure the correlation between the detected photons. (b) Two-photon interference experiment with photons emitted by a single QD under pulsed resonant two-photon excitation, adapted from Ref.\cite{Huber2017}. If the photons are indistinguishable (co-polarized) the height of the central peak is reduced with respect to the peak of cross-polarized photons (fully distinguishable photons).} 
	\label{fig:hom}
\end{figure}

The photon indistinguishability can be verified by measuring the two-photon interference (TPI) visibility via the Hong-Ou-Mandel effect \cite{Ou:PhysRev1988}. The concept is presented in Fig. \ref{fig:hom} (a). Two photons impinge simultaneously on a 50/50 beam-splitter. If the two photons are indistinguishable in all degrees of freedom (energy, polarization, temporal- and spatial overlap), they will leave the beam splitter always through the same output port. Hence, a measured histogram will show no coincidence counts between the detectors at zero time delay.  

The two-photon interference visibility of a single source can be measured by the generation of two single photons via two delayed laser pulses and a respective compensation of the excitation time gap by the aid of an unbalanced Mach-Zehnder interferometer \cite{Gazzano:16}. The result of such a measurement is shown in Fig. \ref{fig:hom} (b). The co-polarized (indistinguishable in polarization) photons show a drop in the central peak with respect to the cross-polarized photons (fully distinguishable). The visibility is given by:

\begin{equation}
      V=1-\frac{g_{\|}}{g_{\bot}},
      \label{eq:HOM}
\end{equation} 

where $g_{\|}$ and $g_{\bot}$ is the integrated peak area around zero time delay for co-polarized and cross-polarized, respectively. A visibility equal to one indicates perfect indistinguishability, while $V=0$ means fully distinguishable photons. Bearing in mind the actual applications of these sources in quantum networks \cite{Kimble:Nat2008,northup2014quantum}, the interference visibility of remote, entangled sources is particularly relevant. In the past years several experiments have been performed using remote QDs \cite{PhysRevB.89.035313,thoma2017two,Reindl:nano2017}. The best remote visibility up to date was achieved by Reindl et al. in 2017 with V=0.51(5) using GaAs QDs embedded in a planar DBR cavity under phonon assisted two-photon excitation (see Ref. \cite{Reindl:nano2017}).  

\subsection{Source brightness}

In spite of this progress, a severe limitation towards desired applications remains the difficulties of extracting the entangled photons generated by QDs from the host matrix. However, whereas sources based on parametric down-conversion suffer from an intrinsic trade-off between source brightness and entanglement fidelity \cite{wang2016experimental} due to the probabilistic nature of the generation process, the deterministic QD-based sources could in principle reach a close-to-unity efficiency without negatively influencing the entanglement fidelity. So far this issue has been mainly addressed from the perspective of single-photon sources.

The efficiency of a single-photon source can be classified by its brightness ($B$), which denotes the probability of collecting a single photon with the first lens of the collection optics upon an excitation pulse. Following a common definition \cite{senellart2017high} the brightness $B=p_{s}\eta_{qe}\eta_{ee}$ is composed of the probability that the QD is excited in the desired target state ($p_{s}$), the quantum efficiency accounting for losses due to non-radiative recombination processes ($\eta_{qe}$) and the extraction efficiency ($\eta_{ee}$). 

The low efficiency of QD-based sources arises mainly from a small extraction efficiency. For achieving a proper 3D confinement, QDs have to be capped and are therefore embedded in a host matrix like GaAs or AlGaAs with a high refractive index (e.g. $n(\mathrm{GaAs})\approx3.5$). For planar as-grown samples the extraction efficiency is therefore limited by total internal reflection (TIR), which allows only a small fraction ($1/(4n^2)$) of the emitted light to exit the sample. Taking into account the finite numerical aperture (NA) of the collection optics commonly only less than $2\%$ of the emitted light can be collected from the top surfaces of planar unprocessed samples \cite{Barnes2002}. Over the past years several paths have been followed in order to improve the brightness of QD-based sources whereby the principle strategies are evident from the definition of the extraction efficiency \cite{gazzano2016toward} given by
 
\begin{equation}
\eta_{ee}=\eta_{ce}\beta=\frac{\eta_{ce}\Gamma_{tm}}{\Gamma_{tm}+\Gamma_{om}}
\label{Eq:PEE}
\end{equation}
 
where $\eta_{ce}$ describes the fraction of the emitted light collectible via the objective lens (with a given NA) and the $\beta$-factor accounts for the spontaneous emission (SE) rates in a specific target mode $\Gamma_{tm}$ or into all other modes $\Gamma_{om}$, respectively. Consequently three main approaches to increase the extraction efficiency were followed: (i) enhancement of the SE rate $\Gamma_{tm}$ into a (cavity) target mode exploiting the Purcell effect (cavity approach), (ii) inhibition of the emission into modes that cannot be collected $\Gamma_{om}$ (waveguide approach) or (iii) increase of $\eta_{ce}$ e.g. by levering the TIR limitation (geometrical approach). In the following, we want to review several concepts possibly leading to bright entangled photon sources.

\subsubsection{Cavity approach: Micropillars}

The most extensively used approach to increase the source brightness is to embed the QDs into a planar $\lambda$-cavity either defined by DBRs or metal mirror(s) \cite{Barnes2002}. The mirrors induce a vertical confinement of the light field leading to a local change in the optical mode density. If the QD emission is properly coupled to such a cavity mode a significant acceleration of the SE rate $\Gamma_{tm}$ (quantified by the Purcell factor $F_p$) can be achieved \cite{purcell1946spontaneous}. Furthermore the diminished transition lifetime limits the sensitivity of the QDs to various dephasing mechanisms induced by the interaction with the solid-state environment. Hence, these structures yield excellent values of single-photon purity, two-photon interference visibility and entanglement fidelity. However, the achieved (pair) extraction efficiencies are still modest (see comparison in Ref. \cite{Jons2017}) even if additionally furnished with a solid immersion lens \cite{Huber2018}.

A decisive step towards the development of bright QD single-photon sources exploiting cavity quantum electrodynamics was the establishment of an additional lateral optical confinement by etching micropillars out of planar cavity samples \cite{gerard1998enhanced}. Initially the randomness of self-assembled QDs in terms of spatial position and spectral properties has been a major hurdle to fully exploit the capabilities of this structures as any spectral and spatial mismatch drastically degrade the brightness of the source. The essential breakthrough for this approach has been accomplished in 2008 by A.~Dousse \textit{et al.} \cite{dousse2008controlled}. In the related work a novel far-field optical lithography technique has been demonstrated capable of accurate spatial alignment (50 nm accuracy) between QD location and micropillar. The lithography is performed \textit{in situ} at low temperature and further allows a spectral preselection of the QD and according customization of the pillar dimension to provide optimal spectral matching between QD emission and fundamental cavity mode. Using this technique, extraction efficiencies as high as 79\% have been demonstrated \cite{gazzano2013bright} whereby additional spectral fine tuning is done via temperature. This has been achieved while simultaneously preserving excellent values for single-photon purity $g^{(2)}(0)=0.0028(12)$ \cite{Senellart:NatPhoton2016} and indistinguishability in terms of (corrected) TPI visibility $V=98.5(1)\%$ \cite{PhysRevLett.116.020401}, both measured under strict resonant excitation (excitation power corresponding to $\pi$-pulse) and additional stabilization of the charge environment via electric fields. Up to now QDs embedded in micropillar cavities show the best performance in terms of high brightness, high single-photon purity and high photon indistinguishability and hence are already applied in scalable multi-photon experiments such as boson sampling \cite{loredo2017boson}.

However, these structures are less suitable for entangled photon pair generation, as the energy difference between biexciton and exciton (relative binding energy) typically exceeds the width of the cavity resonance. We note here that the relative binding energy of the biexciton strongly depends on the QD structure \cite{PhysRevB.88.155312} and that QDs with small binding energy can be found by careful preselection \cite{heindel2017bright} or post-growth tuning via electric or strain fields \cite{PhysRevB.88.155312,PhysRevLett.104.067405}. Unless one resorts to the concept of time-reordering \cite{PhysRevLett.100.120501} - which is limited to non-unity values of the entanglement fidelity \cite{PhysRevB.78.155305} - achieving at the same time vanishing FSS, binding energy and mode matching is however challenging. In addition resonant two-photon excitation becomes impracticable under these conditions. In 2010 Dousse \textit{et al}. \cite{Dousse:Nat2010} presented the concept of a photonic molecule, using two micropillar cavities coupled to a single QD. The authors achieved an entanglement fidelity of $f^{+}=0.68$ at a photon-pair extraction efficiency of $12\%$. The entanglement was essentially limited by the FSS ($<3$ $\mu$eV) and recapture processes, as pulsed, non-resonant excitation was applied. The two-photon interference visibility or single photon purity were not addressed in this work. However, the authors claim that the fine structure could be tuned via electrical field and resonant excitation could be in principle implemented. Unfortunately, no additional improvements in terms of entanglement on this structure exist (a fact that probably arise from the complexity of its fabrication). However, this is to date the the brightest known entangled photon source using a QD emitter \cite{Jons2017}. Since it is a challenging task to use narrowband cavity designs for entangled photon pair extraction, we foresee that cavity structures featuring broadband extraction-efficiency enhancement and modest Purcell factor are more suitable for the purpose.

In this context, L.~Sapienza \textit{et al}. \cite{sapienza2015nanoscale} recently presented a promising alternative approach based on circular Bragg grating 'bullseye' cavities. Hereby, a two-color photoluminescence imaging technique has been used to precisely align the customized cavities to preselected QDs. Sizable extraction efficiencies of up to $48.5\%$ (0.4 NA) over a moderate spectral bandwidth of a few nm have been obtained whereby calculations indicate possible values of up to $80\%$ when using objectives with higher NA. Furthermore, autocorrelation measurements performed under pulsed quasi-resonant excitation yielded an excellent single photon purity of $99.1\%$ ($g^{(2)}(0)<0.009(5)$).  

\subsubsection{Waveguiding approach: Nanowires }

\begin{figure}[htbp]
	\centering
		\includegraphics[width=80mm]{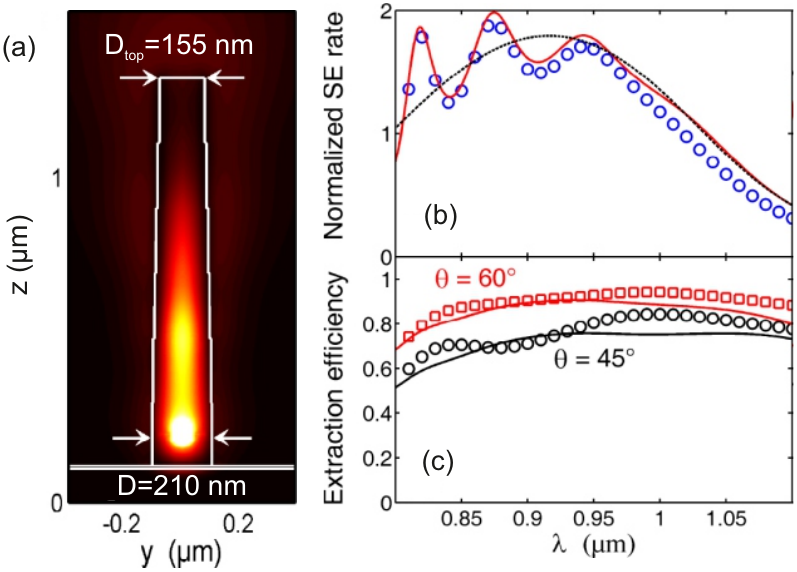}
	\caption{\textbf{Nanowire waveguide.} (a) Color-coded intensity distribution, (b) (normalized) SE rate and (c) extraction efficiency for a QD embedded in a NW WG. Hereby, an optimal coupling to the fundamental WG mode ($\mathrm{HE_{11}}$) has been obtained for a ratio $D/\lambda\approx0.22$ between the waveguide thickness ($D$) at the QD position and the emission wavelength ($\lambda$) supporting large $\beta$-factors $>0.95$. To achieve high extraction efficiencies the NW WG features a planar Ag bottom mirror coated with a thin silica dielectric layer to suppress plasmonic effects at the metal/semiconductor interface. Furthermore, a taper with a small angle of around $1.5^{\circ}$ towards the NW tip is used to reduce scattering of the guided mode, increasing transmission at the top facet and reducing the beam divergence. The QD is located at the NW axis in an anti-node of the electric field to benefit from constructive interference. The calculations of the spontaneous emission rate (b) have been perform using different theoretical models (see Ref.~\cite{friedler2009solid}) and the values obtained for the extraction efficiency are presented for two different NAs corresponding to collection angles of $\theta=60^{\circ}$ (NA 0.87) and $\theta=45^{\circ}$ (NA 0.7). Figures adapted from Ref.~\cite{friedler2009solid}.} 
	\label{nanowires}
\end{figure}

A superior concept for a single photon source which promises extraction efficiencies even over 90\% along a $70\,\mathrm{nm}$-wide spectral range has been proposed in 2009 by I.~Friedler \textit{et al.} \cite{friedler2009solid}. Here a single QD is integrated in a vertical tapered semiconductor nanowire (NW) waveguide, where, instead of using the Purcell effect for enhancement of the spontaneous emission into a cavity mode, the coupling over all modes except of the fundamental waveguide (WG) mode is suppressed. The obtained optimal device layout is shown in Fig.~\ref{nanowires} along with the calculated SE rates and extraction efficiencies for variable emitter wavelengths. The first experimental realization of such a device was pioneered by J.~Claudon \textit{et al.} \cite{claudon2010highly} in 2010. In this work, $2.5\,\mathrm{\mu m}$ high and $200\,\mathrm{nm}$ thick NWs were engineered from a GaAs wafer hosting (high-density) self-assembled In(Ga)As QDs using a top-down approach for fabrication. Auto-correlation measurements of the saturated exciton transition using a HBT setup revealed a very pure single-photon emission of $g^{(2)}(0)<0.008$ for pulsed non-resonant excitation. For the presented device design an excellent extraction efficiency of $\eta_{ee}=72(9)\%$ (using a 0.75 NA) has been obtained. The extraction efficiency was found to be strongly NA-dependent indicating that the far-field emission pattern is not entirely collected. The discrepancy of the obtained source efficiency compared to the theoretically predicted values is attributed to the sensitivity to deviations from the optimal taper angle. This issue has been addressed in a subsequent work dealing with a novel device called "photonic trumpet" \cite{munsch2013dielectric}. The revised design is more tolerant against deviations from the optimal taper angle, provides a higher directionality of the far field emission and shows an improved extraction efficiency of $75(10)\%$.

Although the radial alignment of the QD with respect to the NW axis has been found to demand for comparable low position accuracy \cite{bleuse2011inhibition}, a more accurate alignment between the QD and the photonic structure is required for a further improvement of the source brightness and fabrication yield. This could be either achieved by employing deterministic processing technologies \cite{dousse2008controlled} or using a bottom-up growth approach for the NWs. In the latter case especially wurtzite In(As)P QDs in [111]-oriented (tapered) InP NWs have been studied extensively. This approach turned out to be highly promising for a bright source of polarization-entangled photon-pairs without the necessity of post-growth engineering of the QD properties or temporal post selection, a major source for photon losses. 

These NWs are actually grown using a combination of selective-area and vapor-liquid-solid (VLS) epitaxy \cite{dalacu2011selective}. Hereby the (111)B InP substrate is initially coated with a $\mathrm{SiO_2}$ film. Circular apertures are created in the oxide at predefined positions by means of electron-beam lithography (EBL) and (isotropic) under-etching of the resist pattern with hydrofluoric (HF) acid. Subsequently, the Au catalysts for VLS growth of the NW core are deposited in the center of the oxide openings and the resist is removed. As VLS growth occurs exclusively at the Au/InP interface, the lateral dimensions of the NW core can be controlled via the hole sizes of the EBL pattern ($\pm 2\,\mathrm{nm}$ precision). After growth of the InP NW core containing a thin In(As)P segment, axial growth is suppressed by increasing the temperature. This in turn supports the radial growth of the NW shell which confines the In(As)P QD and defines the WG. As the radial growth of the NW shell is limited to the diameter of the oxide openings \cite{dalacu2009selective} both, lateral QD and WG dimensions can be controlled independently whereby the QD is exactly aligned on the NW axis. This enables an optimal device design and coupling between QD dipole and guided mode ensuring directional emission and efficient light extraction.

In 2012 M.~E.~Reimer \textit{et al.} \cite{reimer2012bright} presented a single-photon source based on a bottom-up grown InP NW with embedded In(As)P QDs. As growth on a metallic mirror is not possible, the NWs have been harvested using a flexible and fully transparent polydimethylsiloxane (PDMS) polymer film followed by evaporation of Au mirror at the bottom side of the NWs. For evaluating the extraction efficiency, PL studies under pulsed non-resonant excitation have been performed. At saturation an extraction efficiency of $42\%$ is obtained for a 0.75 NA, whereby the moderate value has been mainly attributed to a low modal reflectivity of the Au mirror of only 30\%.

Although growth was performed on a (111) substrate a FSS of around $30\,\mathrm{\mu eV}$ was observed for the measured QDs. However, subsequent improvements in the crystal quality of the NWs \cite{dalacu2012ultraclean} and optimization of the QD dimensions finally allowed the demonstration of polarization-entangled photon-pairs in 2014 by T.~Huber \textit{et al.} \cite{huber2014polarization} and M.~A.~M.~Versteegh \textit{et al.} \cite{versteegh2014observation}. In the latter work half of the measured QDs had a very low FSS of $<2\,\mathrm{\mu eV}$. Quantum state tomography revealed a fidelity to the entangled state $(\ket{JJ}+\ket{WW})/\sqrt{2}$ ($J$ and $W$ are orthogonal elliptic polarizations) of $f=0.76(2)$ for the full time window of $6.02\,\mathrm{ns}$. Applying temporal post selection the value could be improved to a fidelity of $f=0.85(6)$ for the narrowest time window of $0.13\,\mathrm{ns}$. The observation of the $(\ket{JJ}+\ket{WW})/\sqrt{2}$ two-photon state instead of the common, maximally entangled $(\ket{HH}+\ket{VV})/\sqrt{2}$ has been attributed to NW anisotropy (elongated cross-section) formed during shell growth (not related to QD shape) giving rise to birefringence in the NW WG, rotating the polarization state of the emitted photon pairs. Using quasi-resonant excitation scheme K.~D.~J\"ons \textit{et al.} \cite{jons2017bright} could improve the fidelity further up to $f^+=0.82(2)$ ($f^+=0.85(9)$) without (with) temporal post-selection using time windows of $4.48\,\mathrm{ns}$ ($0.13\,\mathrm{ns}$).

So far bottom-up grown NW WG with embedded QDs represent the best compromise between source brightness and achieved entanglement fidelity. The weak-point of the NW approach remains the limited visibility in TPI experiments \cite{Reimer012016} despite sub-Kelvin temperatures, an issue which may be solved by resonant excitation.

\subsubsection{Geometrical approach: Microlenses}

\begin{figure}[htbp]
	\centering
		\includegraphics[width=80mm]{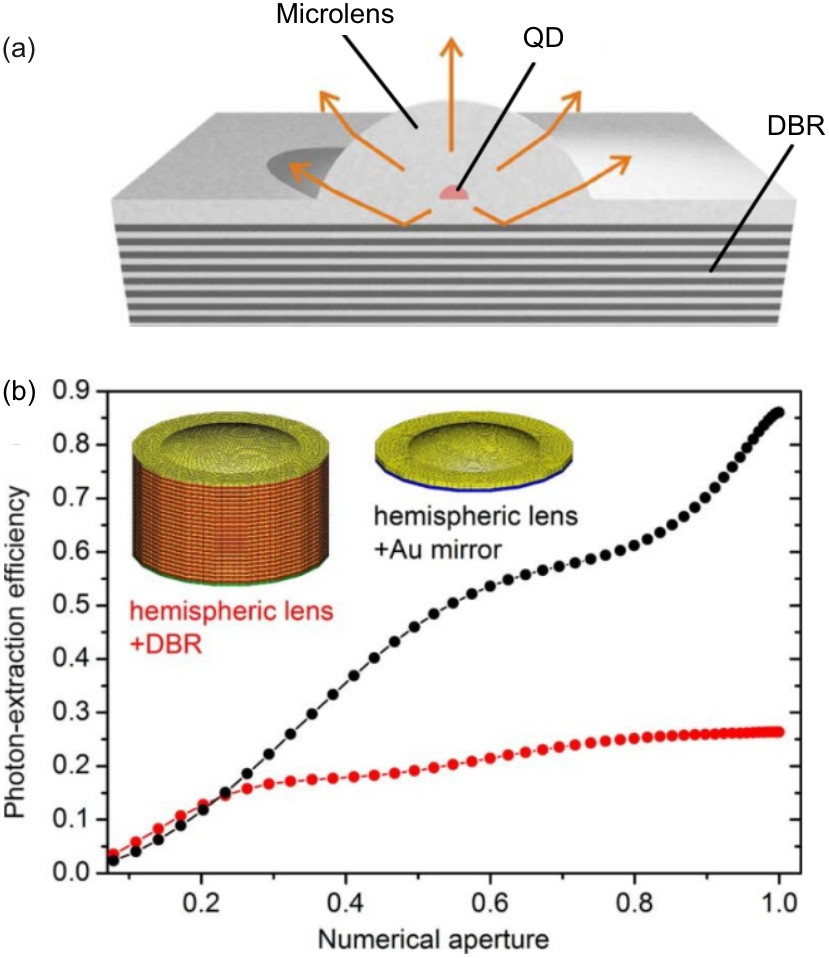}
	\caption{\textbf{Quantum dot microlenses.} (a) Illustration of a single QD grown above a bottom DBR and deterministically embedded in a monolithic microlens allowing for an enhanced photon extraction efficiency [adapted from \cite{heindel2017bright}]. (b) Comparison of the numerically calculated extraction efficiency versus the numerical aperture (NA) of the collection optics for the use of a bottom DBR (red) or a gold mirror (black). In both cases a hemispheric-section lens shape has been assumed but the structure dimensions have been optimized for the respective mirror (black: height $400\,\mathrm{nm}$, base width $2\,\mathrm{\mu m}$, gold mirror $150\,\mathrm{nm}$ below the QD; red: height $400\,\mathrm{nm}$, base width $2.4\,\mathrm{\mu m}$ with a DBR $65\,\mathrm{nm}$ below the QD). For large NAs the calculations reveal that the design with Au-mirror outperforms the DBR as a consequence of the DBRs finite acceptance angle of $<20^{\circ}$ whereas the reflectivity of the Au-mirror is almost independent of the angle of incidence [adapted from \cite{gschrey2015highly}].} 
	\label{lenses}
\end{figure}

A geometrical approach to increase the TIR-limited extraction efficiency via shaping the sample surface at the emitter position has been reported by M.~Gschrey \textit{et al.} \cite{gschrey2015highly} in 2015. In the presented work, monolithic microlenses have been processed deterministically onto self-assembled In(Ga)As QDs grown above a 23-pair $\mathrm{Al_{0.9}Ga_{0.1}As/GaAs}$ DBR [see Fig.~\ref{lenses}(a)]. Spectral pre-selection of the QDs as well as accurate positioning and dimensioning of the microlenses have been achieved using a novel technique called low-temperature cathodoluminescence lithography (CLL) \cite{gschrey2013situ}, which represents a combination of cathodoluminescence (CL) spectroscopy and \textit{in situ} 3D EBL, both performed at liquid He temperature. Due to the weak directionality of the emission, the extraction efficiency achievable with microlenses strongly depends on the NA of the used collection system [see Fig.~\ref{lenses}(d)]. Up to now, using this device concept, only moderate values of up to $\eta_{ee}=29(3)\%$ \cite{schlehahn2015single} could be obtained whereby most reports refer to a relatively low $\mathrm{NA}=0.4$. However, the improvement of the extraction efficiency has been obtained while maintaining ideal values of single-photon purity ($g^{(2)}(0)\leq 0.01$). Furthermore, two-photon interference experiments yielded competitive visibility values of up to $V=94(6)\%$ \cite{thoma2016exploring} using a quasi-resonant two-pulse excitation schemes with a pulse separation of $2\,\mathrm{ns}$. Both values indicate that the lens processing has no drawback on the quantum properties (i.e single photon purity and indistinguishability) of light emitted by the embedded QD. The enabled spectral pre-selection of QDs furthermore facilitates the application in experiments with spatially separated sources \cite{thoma2017two}.\\
Various paths have been followed to achieve a further increase of the extraction efficiency using the microlens approach. Recently, a corresponding value of $\eta=40(4)\%$ has been reported \cite{fischbach2017single} for the remarkable combination of a deterministic QD microlens with a 3D-printed multi-lens micro-objective \cite{gissibl2016two} (with an $\mathrm{NA}\approx0.7$), a concept which is especially interesting for coupling to low-NA fibers. For high-NA collection systems anti-reflection coatings \cite{schnauber2015bright} and in particular the use of an Au-mirror instead of the bottom DBR \cite{fischbach2017efficient} promise a significant increase of the practically achievable extraction efficiency [see Fig.~\ref{lenses}(d)]. In the latter case, the reported application of a flip-chip process easily enables integration with existing strain-tuning technology \cite{martin2017strain} allowing for post-growth engineering of various QD properties \cite{aberl2017inversion}, in particular a cancellation of the FSS necessary for the generation of strongly entangled photon pairs. In this regard, although a demonstration of entangled photon pair emission is still pending, microlenses provide a highly promising alternative due to their broadband enhancement with a bandwidth of about $50\,\mathrm{nm}$ \cite{jakubczyk2016impact}, technological compatibility \cite{schlehahn2016generating} and flexibility with respect to application to various QD systems.

\subsection{Sources at the telecommunication wavelengths}

Sources compatible with the existing fiber-based photonic infrastructure would strongly facilitate the widespread use of quantum communication. Therefore, it is important to develop entangled photon sources operating within the low loss regime of silica fibers. The ongoing progress in fabrication of QD based entanglement-emitters matching the telecom wavelength (1260-1625 nm) started with first demonstrations in 2005 of QD light emission in the C-band by Miyazawa et al. using InAs/InP QDs  \cite{1347-4065-44-5L-L620} and in the O-Band by Alloing et al. with InAs/GaAs QDs \cite{doi:10.1063/1.1872213}. In the following we summarize the results of some recent works. In 2016 Miyazawa showed high single-photon purity ($g^{(2)}(0)=4.4(2)\times 10^{-4}$) for InAs/InP QDs emitting at 1.5 $\mu$m under quasi-resonant excitation \cite{doi:10.1063/1.4961888}. Further, in 2016 Kim et al. studied two photon interference with an InAs/InP QD embedded in a nanophotonic cavity. The resulting visibility is 18$\%$ (67$\%$) without (with) post selection at a pulse separation of 5 ns under non-resonant excitation \cite{Kim:16}. Although the visibility is significantly lower than reported for non-telecom wavelength QDs \cite{Senellart:NatPhoton2016,PhysRevLett.116.020401}, the authors achieved an impressive outcoupling efficiency $>36\%$, which is among the highest ever observed at this emission wavelength. The fabrication of entanglement ready QDs at telecom wavelength is a challenging task as non-zero FSS values are not easy to achieve. In principle InAs/InP QDs are proposed to have an intrinsic FSS one order of magnitude lower than standard InGa/GaAs QDs \cite{PhysRevLett.101.157405}. Beside this, Liu et al. showed in 2014 that InAs/InAlAs droplet QDs grown on (111)A surfaces combine the advantages of a broad emission spectra covering the O,C and L telecom bands with small FSS $<4$ $\mu$eV (on average 25 $\mu$eV), but the authors show no entanglement measurements \cite{PhysRevB.90.081301}. In 2017 Olbrich et al. presented QDs emitting in the C-Band where a fraction of one third has a FSS below 5 $\mu$eV \cite{doi:10.1063/1.4994145}. The authors used InAs quantum dots embedded in InGaAs barriers and shifted the emission by inserting an additional InGaAs metamorphic buffer. Entanglement measurements revealed a fidelity of 0.61(7) at a FSS of approximately 6.20(7) $\mu$eV. In 2017 Skiba-Szymanska et al. presented a different growth strategy to realize low FSS QDs - emitting in the telecom C-band - by droplet epitaxy of InAs on (001) InP substrate \cite{PhysRevApplied.8.014013} via metalorganic vapour phase epitaxy. The authors claim a mean FSS 4x smaller than for Stranski Krastanow QDs. Furthermore, M\"uller et al. presented in 2018 a quantum light-emitting diode based on InP QDs emitting at 1550 nm \cite{Muller2018}. The authors show a peak entanglement fidelity of 0.87(4) by measuring its time evolution (the time evolution of the fidelity is indicated in Fig. \ref{fig:fig1} (d)). Further, the authors observed entangled photon generation up to device temperatures of 93 K, which is an advantage compared to standard InAs/GaAs and GaAs/AlGaAs QDs typically working at temperatures $<10$ K. Beside this, Huwer et al. presented in 2017 a quantum relay based on a QD emitting in the O-band combined with an O-band diode laser for encoding the polarization states (for details on the relay scheme we refer the interested reader to Ref. \cite{PhysRevApplied.8.024007}). The used QD has a FSS of 9.05(1) $\mu$eV and gives a peak fidelity of 0.920(2) (0.963(3)) to a maximally entangled Bell state (the exact time-evolving state), which is the highest value ever presented under these conditions at telecom wavelength. The quantum relay itself achieves an overall fidelity of 0.95(2). In 2018 H\"ofer et al. integrated InGaAs/GaAs quantum dots emitting at the telecom O-band on an uniaxial piezoelectric actuator. This allowed the authors to tune the FSS of the QDs - with an avarage value of 10(6) $\mu$eV - to values below the setup resolution ($<2$ $\mu$eV) \cite{Hofer2018}. However, the authors did not report on the achievable entanglement. Considering the rapid progress on telecom wavelength QDs, we think that the combination of these dots with tuning techniques presented in Sec. \ref{sec:energytuneable}  may soon enable the realization of highly entangled photons sources compatible with existing fiber networks.  

\section{Outlook}

Polarization entangled photon pairs are now seen as the fundamental building block in quantum information sciences and especially for quantum communication. It was long time believed that there is no real application of quantum entanglement to be harnessed, alongside fundamental research, until the first proposals \cite{bennett1984,ekert1991} to prevent eavesdropping of information transfer utilizing Bell's theorem. From this point, an explosion of technological concepts emerged to distribute entangled photons used as message encrypting keys. The basic underlying mechanism is the already mentioned quantum teleportation \cite{Bennett1993}, providing an instantaneous transfer of quantum information from one location to another without violating the no-cloning theorem \cite{wootters1982}. It ultimately allows to relocate entangled photon pairs at arbitrary distances in terms of the teleportation of mixed states involving independent entanglement resources, also known as entanglement swapping, and requires the efficient use of quantum repeater schemes \cite{Briegel1998,Duan2001,Riedmatten2004,Childress2006,Simon2007}. In view of such quantum relays exploiting QD polarization entanglement, experimental realizations are yet limited but promising. Preliminary investigations of the teleportation of a single photon polarization state was implemented on an entangled light-emitting diode \cite{Nilsson2013}, consequently expanded within a laser-heralded teleportation protocol \cite{Stevenson2013}. Remarkably, such a protocol relying on laser generated input states was additionally carried out at telecom wavelength \cite{Stevenson2013}, making it potentially compatible to the existing global communication infrastructure, thereby supplying the necessary scalability. A major hurdle to be considered in this context is the high vulnerability of single photon polarization states to environmental dephasing effects, particularly pronounced in long-distance networks. A robust and feasible solution to account for the polarization degradation was established on QDs via time-bin entanglement \cite{Kauten2014}. Here, the QD XX-X cascade is excited in a coherent superposition of two laser pulses arriving subsequently in time. The photons  then pass an equally unbalanced interferometer with respect to the excitation pulses. The information of the creation time is canceled and results in an entangled-state encoded in the respective time-bins. Indeed, it is of practical advantage to have polarization entanglement at hand any time to perform unitary transformations, as required for complex communication networks and solely use time-bin entanglement for faithful, decoherence-free propagation between quantum relay nodes. In fact, polarization-time-bin interfaces allowing for arbitrary conversion among the different quantum superpositions are possible for QDs without loss of entanglement fidelity \cite{Versteegh2015} by creating time-bins, yet after entangled photon pair emission, based on unbalanced, polarization-selective Mach-Zehnder interferometers. In summary, QDs as sources of polarization entangled photons pairs are on the verge to unfold their full potential with respect to actual, practical applications. Once all individually achieved results on QDs are finally merged into a single, semiconductor-based entangled photon-pair source, it will pave the way towards sophisticated and fully deterministic quantum networks.

\newpage

\bibliography{references}

\end{document}